# Critical look at the atmospheric Cu fire-through dielectric metallization for cost-effective and high efficiency silicon solar cells

Donald Intal[1], Sandra Huneycutt[1], Abasifreke Ebong[1], Ajeet Rohatgi[2], Vijay Upadhyaya[2], Sagnik Dasgupta[2], Ruohan Zhong[2], Thad Druffel[3], Ruvini Dharmadasa[3]

[1]University of North Carolina at Charlotte, Charlotte, NC 28223, USA

[2]Georgia Institute of Technology, Atlanta, Georgia 30332, USA

[3]Bert Thin Films LLC, Louisville, KY 40208, USA

*Abstract*—The formation of stable copper–silicide ($Cu_3Si$) interfaces is crucial for cost-effective, high-efficiency solar cells. However, copper's diffusivity and electromigration issues pose challenges for contact stability. This study employs Laser-Enhanced Contact Optimization (LECO) to induce localized nano-scale Joule heating at the Cu–Si interface in phosphorus-doped p-PERC solar cells. High-resolution STEM and bright field analyses confirm stable $Cu_3Si$ formation in LECO-treated samples, with significantly reduced material segregation compared to nonLECO samples. SEM and post-etch EDS mapping demonstrate improved chemical resistance and interface cleanliness. Electrically, LECO treatment reduces series resistance by a factor of 3, enhancing fill factor and efficiency while preserving diode quality. These results highlight LECO as a scalable method for reliable, silver-free solar cell metallization.

## I. Introduction

Laser-Enhanced Contact Optimization (LECO) offers an alternative strategy by applying localized, high-density Joule heating directly at the metal–semiconductor interface [1], [2], [3]. LECO selectively activates interface regions without globally heating the entire wafer, thereby preserving the sensitive surface passivation and underlying emitter structures [2], [4]. LECO has demonstrated the ability to enhance Ag–Si bonding by promoting sintering and decreasing contact resistivity [1], [3]. For the Ag-Si contact metallization, this has translated into lower series resistance and improved mechanical adhesion[3], [5] and hence higher fill factor and efficiency boost. Despite the improved efficiency of the solar cells due to the LECO post anneal process, the cost of Ag metallization is still high and should be replaced with an alternative metal such as Cu [6], [7], as highlighted by the long-term divergence in Ag and Cu market prices shown in Figure 1. The extension of LECO to Cu, introduces new interfacial dynamics due to copper's unique chemical behavior [8].

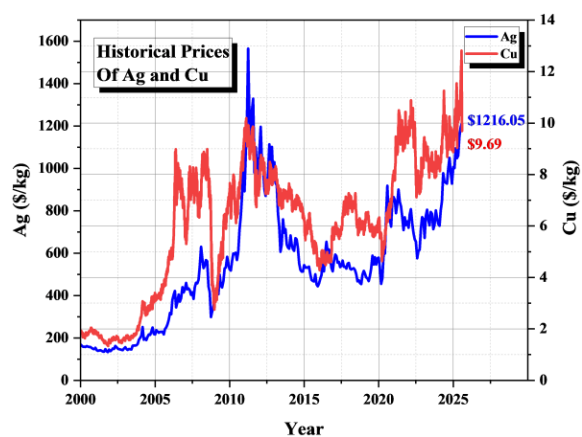

*Figure 1: Historical Ag and Cu market prices over the past two decades [9], [10].*

However, a particular concern in Cu metallization is the formation of hillocks stress-induced protrusions that develop during post-deposition thermal processing [11], [12]. These hillocks can breach dielectric caps or interlayer dielectrics, leading to short circuits and early failure of devices [11], [13]. Traditional approaches to mitigate hillock formation, such as introducing thick dielectric layers, pre-annealing strategies, or doping Cu with aluminum, often require additional process complexity or sacrifice scalability [12], [14], [15]. Likewise, while conventional diffusion barriers like TaN or TiN are widely used to confine Cu, they can introduce high interfacial resistance or complicate integration with low-temperature processes in solar cell manufacturing [13], [16].

For copper (Cu) to be used as a front-side metallization material in solar cells for the next-generation photovoltaics, the challenges associated with high diffusivity, chemical reactivity, and mechanical instability under thermal and electrical stress should be investigated [17], [18], [19], [20],

[21]. These factors give rise to undesirable effects such as contact degradation, diffusion into passivation layers, and electromigration, all of which compromise long-term device performance and yield [22], [23], [24].

Unlike Ag, Cu readily reacts with silicon at moderate temperatures to form a series of intermetallic silicides, notably $Cu_5Si$, $Cu_{15}Si_4$, and $Cu_3Si$, depending on the local silicon availability and processing temperature [25]. Of these, $Cu_3Si$ has received particular attention due to its stability and self-limiting nature during solid-state diffusion. The formation of $Cu_3Si$ at the copper–silicon interface has been shown to reduce the availability of mobile Cu atoms, which are otherwise prone to diffusive and electromigration-driven transport. Thermodynamic analyses have confirmed that this silicidation reaction is mildly exothermic, and kinetic studies show that it proceeds via diffusion-limited mechanisms at temperatures as low as 200–300 °C [25], [26]. Notably, epitaxial growth of $Cu_3Si$ on Cu has been observed under controlled silane treatments, forming coherent and defect-minimized interfaces [27], [28].

In microelectronic applications, copper silicide has been investigated as a passivation material that simultaneously serves as a diffusion barrier and adhesion promoter [29], [30]. The silicide layer "locks" copper atoms into a chemically stable lattice, limiting their mobility and thereby suppressing electromigration and hillock formation. Patent disclosures have further emphasized its utility in reducing oxidation, enhancing barrier performance, and improving electromigration lifetime in damascene interconnect structures [14], [26].

These insights suggest that LECO, by enabling localized thermal activation, may trigger the formation of such interfacial silicide layers in Cu metallization schemes without the need for conventional high-temperature furnace annealing or plasma-enhanced silane environments. However, the interfacial chemistry and structural consequences of LECO-induced silicidation in Cu contacts remain largely unexplored, particularly in the context of solar cell metallization.

To address this gap, we apply LECO to Cu contacts on phosphorus-doped emitters and systematically investigate the resulting interfacial transformations. Using high-resolution Scanning Transmission Electron Microscopy (STEM) and post-etch Scanning Electron Microscopy (SEM), we analyze the structural and compositional changes induced by LECO. Particular attention is given to the emergence of copper silicide phases, the suppression of Cu diffusion, and the potential implications for contact stability and performance.

## II. Experimental Method

Phosphorus-doped p-PERC silicon solar cells were fabricated and metallized with screen-printed Cu front contacts. After metallization, a subset of samples underwent Laser-Enhanced Contact Optimization (LECO), in which localized Joule heating was induced at the metal–semiconductor interface to promote contact formation while avoiding global wafer heating. nonLECO samples were retained as controls for all comparisons. Current–voltage (I–V) measurements were performed before and after LECO to quantify changes in device performance metrics, including series resistance, fill factor, and efficiency.

Interfacial structure was examined using cross-sectional scanning transmission electron microscopy (STEM) in bright-field (BF) mode. To assess chemical durability and residual copper behavior, metallized regions were sequentially etched in 70% $HNO_3$ followed by 2.5% HF, and the post-etch morphology and composition were evaluated by scanning electron microscopy (SEM) at representative regions.

## III. Results and Discussions

Comparative I–V measurements, along with device fabrication, screen printing, and post-metallization LECO processing, were conducted at Georgia Tech University Center for Photovoltaics and Education (UCEP). The data confirms that LECO process significantly enhances the performance of silicon solar cells with Cu-screen-printed contact. Table 1 shows that after LECO treatment, series resistance reduced by a factor of 3, going from 1.9 $\Omega$-$cm^2$ to only 0.7 $\Omega$-$cm^2$ resulting in the final fill factor of 76.2%.

*Table 1: Comparative IV Measurements of screen-printed solar cells before and after LECO treatment.*

| Parameter | nonLECO | LECO |
|---|---|---|
| $V_{oc}$ (V) | 0.664 | 0.675 |
| $J_{SC}$ (mA/$cm^2$) | 39.8 | 39.6 |
| FF (%) | 67.9 | 76.2 |
| Efficiency (%) | 17.9 | 20.4 |
| $R_s$ ($\Omega$-$cm^2$) | 1.9 | 0.7 |
| pFF (%) | 77.4 | 80.0 |
| $J_{o2}$ (A/$cm^2$) | 4.6E-08 | 2.4E-08 |

However, the pseudo fill factor (pFF) improve from 77.4% to 80% while the reverse saturation current density $J_{o2}$ decreased by a factor or two. The result indicates the diode quality is stable, and recombination losses near the metal–semiconductor junction does not increase after LECO process. The observed electrical enhancements are consistent with prior studies on laser-assisted contact optimization, which have shown that reducing series resistance at the contact interface leads to higher FF and overall device efficiency [31], [32].

The performance improvements are closely tied to interfacial structural and chemical changes [32], discussed in the following sections. Section 2A describes the formation of Cu₃Si via interfacial alloying; Section 2B examines its chemical durability; Section 2C evaluates its role in mitigating electromigration; and Section 2D discusses broader implications for LECO-treated Cu contacts in scalable manufacturing.

### A. Interfacial Alloying and Silicide Formation

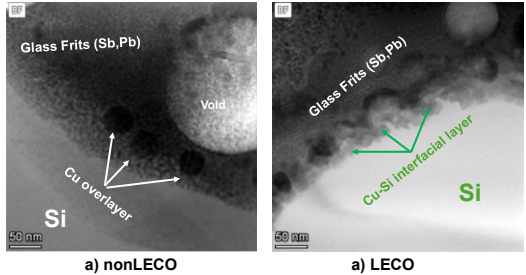

*Figure 2: STEM/EDS elemental mapping of Cu and Si at the contact interface.*

Figure 2 is the cross-sectional STEM bright-field (BF) analysis, which provides direct morphological and elemental evidence of interfacial alloying in LECO-treated Cu-contact sample. EDS mapping revealed localized co-distribution of copper and silicon near the contact interface, indicating the formation of Cu–Si intermetallic phases. The composition and distribution are consistent with the formation of the copper-rich Cu₃Si phase, which is thermodynamically stable and known to emerge under moderate thermal activation. In contrast, nonLECO samples exhibited sharp segregation of copper and silicon, spatially separated by an intact glass frit layer, with no signs of silicide formation.

These observations suggest that the LECO process generates localized thermal conditions sufficient to overcome the activation barrier for solid-state diffusion, allowing silicon atoms to penetrate the copper layer and form silicide. Literature reports confirm that Cu₃Si can form at temperatures as low as 300 °C, with diffusion-limited growth kinetics typical of silicide reactions [25]. The Joule heating delivered by LECO mimics this behavior in a spatially confined region, enabling metallurgical transformation at the Cu–Si interface without subjecting the entire wafer to high-temperature annealing.

### B. Post-Etch Stability and Chemical Resistance

To evaluate interfacial stability, samples were sequentially etched in 70% HNO₃ followed by 2.5% HF. Figure 3 presents SEM/EDS/Spectra analysis of the peak regions, revealing a notable trend: nonLECO samples retained more copper residue than LECO-treated regions, which appeared significantly cleaner post-etched.

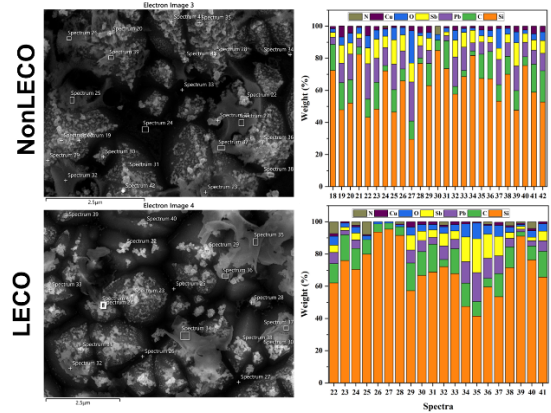

*Figure 3: SEM and EDS analysis at the peak region of LECO and nonLECO solar cells with elemental wt%.*

This observation suggests a key difference in how copper exists in each case. In nonLECO samples, the copper is largely metallic and remains partially embedded in the glass frit or redeposited during etching [33]. While concentrated nitric acid oxidizes Cu to Cu²⁺, the process may be incomplete or locally non-uniform. The redox reaction proceeds as:

$$3Cu(s) + 8HNO_3 \rightarrow 3Cu^{2+}(aq) + 2NO(g) + 4H_2O + 6NO_3^-$$

These released Cu²⁺ ions can react with fluoride from the HF step to form insoluble copper fluoride, which may adhere to the surface and contribute to residual:

$$Cu^{2+}(aq) + 2F^-(aq) \rightarrow CuF_2(s)$$

In addition, the Si substrate can facilitate galvanic displacement, in which Cu²⁺ is reduced and re-deposited as metallic Cu. In the presence of HF, silicon oxidized to form hexafluorosilicic acid ($H_2SiF_6$), while $Cu^{2+}$ is simultaneously reduced to $Cu^0$:

$$Si(s) + 6HF(aq) \rightarrow H_2SiF_6(aq) + 4H^+(aq) + 4e^-$$

$$Cu^{2+}(aq) + 2e^- \rightarrow Cu(s)$$

The net reaction:

$$Si(s) + 6HF(aq) + 2Cu^{2+}(aq) \rightarrow H_2SiF_6(aq) + 4H^+(aq) + 2Cu(s)$$

This galvanic process leads to unintended re-deposition of cu onto the silicon surface, contributing to the EDS-detected residues [34], [35].

In contrast, LECO-treated samples show reduced Cu residue because the Cu has largely reacted to form copper silicide (Cu₃Si). This intermetallic phase is less susceptible to acid dissolution and does not redeposit easily. During etching, any unreacted Cu is removed, and the Cu$_3$Si interface is likely passivated by a thin SiO$_2$ layer formed through partial oxidation [26], [36]. HF may etch this oxide, but the underlying silicide remains thin, adherent, and dispersed yielding a weaker Cu signal in EDS, especially against the strong Si background.

The absence of visible Cu-rich byproducts in LECO-treated regions confirms that localized silicide formation improves the chemical cleanliness and integrity of the contact interface. These findings align with reports of Cu$_3$Si forming a stable, conductive, and corrosion-resistant interfacial layer, reinforcing the role of LECO in both suppressing unwanted residues and enabling high-quality metallization for solar cell manufacturing.

### C. Electromigration Resistance and Hillock Suppression

At the atomic level, Cu$_3$Si formation immobilizes copper atoms within an intermetallic lattice, significantly reducing atomic mobility [37]. This structural stabilization effectively suppresses electromigration, a known reliability failure mode in copper conductors and inhibits hillock formation during thermal cycling. These reliability benefits are well-documented in dual-damascene interconnects and metal–insulator–metal (MIM) capacitors, where silicide formation has been shown to extend electromigration lifetime and prevent thermal extrusion defects [14], [30].

Furthermore, the silicide layer naturally eliminates native copper oxides such as CuO and Cu$_2$O, which are typically brittle and poorly adhered. Their removal during silicidation enhances interfacial adhesion and provides a cleaner, more stable surface for subsequent dielectric deposition. This is especially important in PERC solar cells, where poor contact adhesion can lead to delamination or contact degradation during operation or encapsulation.

### D. Implications for Device Performance and Manufacturing

In addition to its chemical and mechanical benefits, Cu$_3$Si retains good electrical conductivity. Its presence at the contact interface improves the contact homogeneity, which helps to lower the contact resistance. By forming a nanometer-thick, conductive, and self-limiting silicide layer, the LECO process addresses both performance and reliability concerns associated with copper metallization. This enables a path forward for alternative to silver metal contacts without compromising efficiency or longevity as a key goal for cost-effective, high-volume solar cell production. Collectively, these findings validate the LECO process as an effective means of engineering the Cu–Si interface through localized silicidation. The resulting copper silicide layer serves as both a diffusion barrier and mechanical stabilizer, reducing copper mobility, enhancing adhesion, and suppressing failure modes such as electromigration and hillock growth.

## IV. Conclusions

Through comparative analyses of LECO-treated and nonLECO samples, we have shown that LECO facilitates the formation of a chemically stable and structurally copper silicide (Cu$_3$Si) layer at the Cu–Si interface. This intermetallic layer resists chemical etching and preserves the electrical and mechanical integrity of the contact region under processing and operational conditions. Microscopic and spectroscopic characterization confirmed the presence of Cu$_3$Si and related silicide phases, which function as self-aligned diffusion barriers that mitigate copper migration and electromigration, persistent challenges in copper-based metallization. Additionally, the silicide layer improves interfacial adhesion and suppresses hillock and void formation, both of which are critical to ensuring long-term reliability. These findings align with established thermodynamic and kinetic models of silicide formation and are further supported by atomic-scale analyses and prior patent literature.